\documentclass[aip,reprint,onecolumn,jmp,graphicx]{revtex4-1}

\draft 

\usepackage{amsmath}
\usepackage{amssymb}
\usepackage{bm}

\newcommand{\rmd}{\mathrm{d}}
\newcommand{\rme}{\mathrm{e}}
\newcommand{\rmi}{\mathrm{i}}

\begin{document}


\title{Algebraic probability, classical stochastic processes, and counting statistics} 



\author{Jun Ohkubo}
\email[Email address: ]{ohkubo@i.kyoto-u.ac.jp}
\affiliation{
Graduate School of Informatics, Kyoto University,\\
Yoshida Hon-machi, Sakyo-ku, Kyoto-shi, Kyoto 606-8501, Japan
}



\begin{abstract}
We study a connection between the algebraic probability and classical stochastic processes
described by master equations.
Introducing a definition of a state which has not been used for quantum cases,
the classical stochastic processes can be reformulated 
in terms of the algebraic probability.
This reformulation immediately gives the Doi-Peliti formalism,
which has been frequently used in nonequilibrium physics.
As an application of the reformulation,
we give a derivation of basic equations for counting statistics,
which plays an important role in nonequilibrium physics.
\end{abstract}

\pacs{}

\maketitle 

\section{Introduction}
\label{sec_introduction}

A `classical' stochastic process plays an important role
in studies of nonequilibrium physics.
Here, `classical' means that there is no quantum effect.
However, in spite of the `classical' features,
there are many analytical methods analogous to those in quantum mechanics.
One of famous examples is the Doi-Peliti formalism \cite{Doi1976,Doi1976a,Peliti1985},
which is also called the field-theoretic method or second quantization method.
In the Doi-Peliti formalism, bosonic creation and annihilation operators are used,
and due to the similarity with the quantum mechanics,
perturbation calculations and renormalization group methods
have been employed,
which give us useful understandings for nonequilibrium behaviors \cite{Tauber2005}.
Another example of applications of quantum mechanical formulations
to classical stochastic processes is a geometric phase concept.
The geometric phase concept has been used 
in order to discuss nonequilibrium behaviors caused by external perturbations
to stochastic processes \cite{Sinitsyn2007,Ohkubo2007,Ohkubo2008,Sinitsyn2009}.
Discussions for the Berry phase or Aharanov-Anandan phase in quantum mechanics
have been applied, and various analytical results have been obtained.
It is expected that further investigations
for the similarity between quantum mechanical formulations
and the classical stochastic processes
enable us to give useful methods and deep understanding
for nonequilibrium physics.

In the present paper,
we show that a scheme of the algebraic probability gives
a unified way to discuss certain types of classical stochastic processes.
Employing a Jacobi sequence and an interacting Fock space,
we define a state suitable for the classical stochastic processes.
This formulation gives an easy way
to see the mathematical structure of the Doi-Peliti formalism;
a natural connection between the Doi-Peliti formalism and orthogonal polynomials is obtained.
The Doi-Peliti formalism corresponds to 
the canonical commutation relation,
and we will show that the unifying method based on the interacting Fock space
is also applicable for other cases, for example, the canonical anticommutation relations;
as an example, a simple hopping model used in nonequilibrium physics
is described in terms of the algebraic probability.
In addition, as an application of the formulation to other nonequilibrium physics,
we discuss the counting statistics,
which is an important method in order to investigate
nonequilibrium behaviors.
We note that the counting statistics has been used
as a basis for the geometric phase discussions in the classical stochastic processes.
Through the discussions,
a conventional generating function method in the previous works
can be reinterpreted in the algebraic probability.
This reinterpretation is useful to construct a modified transition matrix,
which plays a central role in the counting statistics.
We will also show that this scheme
is available even for the counting statistics for continuous values,
which is needed, for example, to investigate entropy productions
in nonequilibrium systems.

The present paper is constructed as follows.
In Sec.~\ref{sec_algebraic_probability},
we briefly review discussions for a Jacobi sequence, interacting Fock space,
and the algebraic probability.
A definition of a state is also proposed here.
Two examples are given in Sec.~\ref{sec_examples},
in which a connection between the scheme of the algebraic probability
and the Doi-Peliti formalism is also discussed.
In Sec.~\ref{sec_cs},
we apply the scheme of the algebraic probability
to the counting statistics.
Two interacting Fock spaces are combined in order to calculate a net flow,
and it is shown that a modified transition matrix,
which is used in order to describe time evolution equations for generating functions,
is simply constructed.
Finally, we give some concluding remarks in Sec.~\ref{sec_conclusions}.

\section{Algebraic probability for classical stochastic processes}
\label{sec_algebraic_probability}

\subsection{Jacobi sequence and interacting Fock space}
\label{sec_Jacobi}

We give brief explanations for some concepts which is needed to define
a state for classical stochastic processes.
Definitions, properties, and explanations here are based on Ref.~\onlinecite{Hora_Obata_book};
for details, see Ref.~\onlinecite{Hora_Obata_book}.

A \textit{Jacobi sequence} is a sequence $\{\omega_n; n = 1,2, \cdots \}$
which satisfies one of the following two conditions:
(i) $\omega_n > 0$ for all $n$;
(ii) there exists a number $m_0 \ge 1$ such that
$\omega_n = 0$ for all $n \ge m_0$ and $\omega_n > 0$ for all $n < m_0$.
Hereafter, we mainly give discussions for case (i);
in order to discuss case (ii), the number $m_0$ should be introduced adequately.

We next introduce an orthonormal basis $\{\tilde{\Phi}_n; n = 0, 1, \cdots \}$
in an infinite-dimensional Hilbert space.
Using the Jacobi sequence, linear operators $B^{\pm}$ are introduced as follows:
\begin{align}
&B^+ \tilde{\Phi}_n = \sqrt{\omega_{n+1}} \tilde{\Phi}_{n+1}, \quad n = 0, 1, \cdots,\\
&B^- \tilde{\Phi}_0 = 0, \qquad B^- \tilde{\Phi}_n = \sqrt{\omega_n} \tilde{\Phi}_{n-1}, \quad n = 1,2, \cdots.
\end{align}
We will call $B^+$ a \textit{creation operator}, and $B^-$ an \textit{annihilation operator}.
The connection among the orthonormal basis $\{\tilde{\Phi}_n\}$, the Jacobi sequence $\{\omega_n\}$
and the linear operators $B^{\pm}$ is expressed as follows schematically:
\begin{align*}
\tilde{\Phi}_0 
\quad
\begin{matrix}
\sqrt{\omega_1} \\ \rightleftarrows \\ \sqrt{\omega_1}
\end{matrix} 
\quad
\tilde{\Phi}_1
\quad
\begin{matrix}
\sqrt{\omega_2} \\ \rightleftarrows \\ \sqrt{\omega_2}
\end{matrix} 
\quad
\tilde{\Phi}_2
\quad
\rightleftarrows
\quad
\cdots
\quad
\rightleftarrows
\quad
\tilde{\Phi}_{n-1}
\quad
\begin{matrix}
\sqrt{\omega_n} \\ \rightleftarrows \\ \sqrt{\omega_n}
\end{matrix} 
\quad 
\tilde{\Phi}_{n}
\quad
\begin{matrix}
\sqrt{\omega_{n+1}} \\ \rightleftarrows \\ \sqrt{\omega_{n+1}}
\end{matrix} 
\quad 
\tilde{\Phi}_{n+1}
\quad \rightleftarrows 
\quad \cdots.
\end{align*}
$\Gamma$ is defined as a linear subspace spanned by
$\{(B^+)^n \tilde{\Phi}_0 ; n = 0,1,2, \dots \}$,
and the inner product of $\Gamma$ is denoted by $\langle \cdot , \cdot \rangle$;
by construction, $\langle \tilde{\Phi}_m, \tilde{\Phi}_n \rangle = \delta_{m,n}$,
where $\delta_{m,n}$ is the Kronecker delta.
The quadruple $\Gamma_{\{\omega_n\}} = (\Gamma, \{\tilde{\Phi}_n\},B^+,B^-)$
is called an \textit{interacting Fock space} associated with the Jacobi 
sequence $\{\omega_n\}$.
In addition to the linear operators $B^{\pm}$,
an additional operator $N$ is defined as
\begin{align}
N \tilde{\Phi}_n = n \tilde{\Phi}_n.
\end{align}
The operator $N$ is called the \textit{number operator},
and it is used frequently in the following discussions.

While we introduce the orthonormal basis $\{\tilde{\Phi}_n\}$,
it is useful to define a slightly different basis, i.e., an orthogonal basis, 
in order to discuss the classical stochastic processes.
We define a basis $\{\Phi_n\}$ as follows:
\begin{align}
\Phi_n \equiv \sqrt{\omega_n \cdots \omega_1} \tilde{\Phi}_n.
\end{align}
Hence, 
\begin{align}
&B^+ \Phi_n = \Phi_{n+1}, \quad n = 0, 1, \cdots,\\
&B^- \Phi_0 = 0, \qquad B^- \Phi_n = \omega_n \Phi_{n-1}, \quad n = 1, 2, \cdots,
\end{align}
and $\langle \Phi_m, \Phi_n \rangle = \omega_n \cdots \omega_1 \delta_{m,n}$.
The schematic expression of the redefined basis is as follows:
\begin{align*}
\Phi_0 
\quad
\begin{matrix}
1 \\ \rightleftarrows \\ \omega_1
\end{matrix} 
\quad
\Phi_1
\quad
\begin{matrix}
1 \\ \rightleftarrows \\ \omega_2
\end{matrix} 
\quad
\Phi_2
\quad
\rightleftarrows
\quad
\cdots
\quad
\rightleftarrows
\quad
\Phi_{n-1}
\quad
\begin{matrix}
1 \\ \rightleftarrows \\ \omega_n
\end{matrix} 
\quad 
\Phi_{n}
\quad
\begin{matrix}
1 \\ \rightleftarrows \\ \omega_{n+1}
\end{matrix} 
\quad 
\Phi_{n+1}
\quad \rightleftarrows 
\quad \cdots.
\end{align*}

As shown in Sec.~\ref{sec_examples}, 
this scheme is useful to deal with the classical stochastic processes,
especially chemical reaction systems or particle creation-annihilation systems.

\subsection{Brief review of algebraic probability}
\label{sec_review}

We here give a brief explanation of the algebraic probability.
The algebraic probability is also called quantum probability or noncommutative probability.
For details, see Ref.~\onlinecite{Hora_Obata_book}.

The algebraic probability is based on a $*$-algebra and a state defined on the $*$-algebra.
In order to define a $*$-algebra, we firstly introduce an \textit{involution},
which is defined as a map $a \mapsto a^*$ on an algebra $\mathcal{A}$
and satisfies
\begin{align*}
(a+b)^* = a^* + b^*, \quad (\lambda a)^* = \bar{\lambda} a^*, \quad
(ab)^* = b^* a^*, \quad (a^*)^* = a
\end{align*}
for $a,b \in \mathcal{A}$ and $\lambda \in \mathbb{C}$.
A \textit{$*$-algebra} is an algebra equipped with an involution.
Next, we define a linear function $\varphi$ on a $*$-algebra $\mathcal{A}$
with values in $\mathbb{C}$, which satisfies the following two properties:
(i) (positivity) $\varphi(a^* a) \ge 0$;
(ii) (normality)  $\varphi(1_\mathcal{A}) = 1$,
where $1_\mathcal{A}$ is the identity.
A linear function $\varphi$ with the positivity and normality is called a \textit{state},
and an \textit{algebraic probability space} is a pair 
$(\mathcal{A},\varphi)$ of a $*$-algebra $\mathcal{A}$ and a state $\varphi$ on it.

\subsection{Proposal of a definition of state for classical stochastic processes}
\label{sec_definition_of_state}


We here consider a stochastic process with discrete states
indexed by $n = 0, 1, 2, \cdots$.
The probability distribution for the stochastic process
is described as $P(n)$, $n = 0,1,2,\cdots$.
For example, the state $n$ corresponds to a number of certain chemical substances
for chemical reaction systems.
It is easy to extend the following scheme to multivariate cases,
so that we explain only a single variable case for simplicity.

One of the main claims in the present paper is a useful definition of state $\varphi$
for the classical stochastic process.
Firstly, we consider a certain Jacobi sequence, which is suitable to deal with
the classical stochastic process.
Some examples for the Jacobi sequences are shown in Sec.~\ref{sec_examples}.
Secondly, in association with the Jacobi sequence, 
an orthogonal basis $\{\Phi_n\}$, the linear operators $B^{\pm}$,
and the number operator $N$ are introduced, explained in Sec.~\ref{sec_Jacobi}.
Thirdly, we consider a $*$-algebra $\mathcal{A}$ constructed by $B^{\pm}$ and $N$.
Note that we define $(B^{\pm})^* = B^{\mp}$ and $N^* = N$.
Finally, a state $\varphi$ on $\mathcal{A}$ is defined as follows:
\begin{align}
\varphi(a) \equiv \left\langle \left( \sum_m \frac{1}{\omega_m \cdots \omega_1}  \Phi_m \right),
a
\left( \sum_{n} P(n) \Phi_n \right)
\right\rangle
= \sum_{m,n} \frac{P(m)}{\omega_m \cdots \omega_1} \langle \Phi_m, a \Phi_n \rangle,
\label{eq_definition_of_state}
\end{align}
for $a \in \mathcal{A}$.
Since $P(n)$ is the probability distribution, 
the positivity and normality of the state $\varphi$ are guaranteed.

In order to explain the reason why we should use $P(n)$, not $\sqrt{P(n)}$ like quantum mechanics,
the following formal expressions are useful.
Firstly, we focus on a time evolution of the classical stochastic process.
The time evolution is governed by a master equation 
\begin{align}
\frac{\rmd }{\rmd t} P(n,t) = \sum_{m} K_{nm} P(m,t),
\label{eq_original_master_equation}
\end{align}
where $P(n,t)$ is the probability with which the system is in state $n$ at time $t$,
and $K = \{K_{nm}\}$ a transition matrix.
(It is also possible to treat a time-dependent transition matrix.)
We set the initial condition as $P(n,0) = P(n)$.
Here, a \textit{ket vector} $| n \rangle$ is formally defined as
\begin{align}
| n \rangle = \Phi_n,
\end{align}
and we introduce a \textit{state vector} $| \psi(t) \rangle$ as follows:
\begin{align}
| \psi(t) \rangle = \sum_n P(n,t) \Phi_n.
\label{eq_state_vector}
\end{align}
Instead of the time evolution of $P(n,t)$,
we deal with the time evolution of the state vector $| \psi(t) \rangle$ as follows:
\begin{align}
\frac{\rmd}{\rmd t} | \psi(t) \rangle = L(B^+,B^-,N) | \psi(t) \rangle,
\label{eq_time_evolution_of_state_vector}
\end{align}
where $L(B^+,B^-,N)$ is a linear operator constructed by $B^{\pm}$ and $N$.
It is possible in general to construct the linear operator $L(B^+,B^-,N)$
in order to recover original master equations \eqref{eq_original_master_equation};
in this case, each coefficient of the state vector $|\psi(t) \rangle$ at time $t$
corresponds to $P(n,t)$ for the original master equation adequately.
The state vector $| \psi(t) \rangle$ at time $t$ is formally written as
\begin{align}
| \psi(t) \rangle =  \exp\left( L(B^+,B^-,N) t \right)  | \psi \rangle,
\end{align}
where $| \psi \rangle = \sum_n P(n) \Phi_n$.
Next, we define a \textit{bra vector} $\langle m |$
as a combination of $\Phi_m$ and the rest of the inner product;
for example, when $\{\Phi_n\}$ are expressed by orthogonal polynomials,
$\langle m |$ is defined as the integral part of the inner product and $\Phi_m$;
please see Sec.~\ref{sec_examples}, 
in which examples of the definitions of the bra vectors $\langle m |$ are shown.
Using the bra vectors $\{\langle m |\}$,
a \textit{projection state vector} is introduced as follows:
\begin{align}
\langle \mathrm{P} | \equiv \sum_m \frac{1}{\omega_n \cdots \omega_1} \langle m |.
\end{align}
Hence, the state $\varphi$ is redefined in terms of the projection state vector
and the state vector:
\begin{align}
\varphi(a) \equiv \langle \mathcal{P} | a | \psi \rangle,
\label{eq_new_definition_of_state}
\end{align}
for $a \in \mathcal{A}$.
This expression using the bra and ket vectors might be familiar with physicists.
Using the above redefinition,
a statistical average of a physical quantity
at time $t$, $a(t)$, is obtained by $\varphi(a(t)) = \varphi( a \exp\left( L(B^+,B^-,N) t \right) )$.
Note that the time evolution operator is also in the $*$-algebra:
$\exp\left( L(B^+,B^-,N) t \right) \in \mathcal{A}$.
From the above constructions,
it is clear that we should not use $\sqrt{P(n)}$ in the definitions;
if we use $\sqrt{P(n)}$,
the time evolution of a state vector could not be connected to the original master equation.

We finally comment on the bra and ket vectors.
These vectors and the operators $\{B^{\pm}\}$ are related to each other as follows:
\begin{align}
&B^+ | n \rangle = | n+1 \rangle, \quad n = 0,1,2,\cdots, \label{eq_op_start}\\
&B^- | 0 \rangle = 0, \qquad B^- | n \rangle = \omega_n | n-1 \rangle, \quad n = 1,2,\cdots,\\
&\langle n | B^- = \langle n+1 |, \quad n = 0,1,2, \cdots, \\
&\langle 0 | B^+ = 0, \qquad \langle n | B^+ = \langle n-1 | \omega_n, \quad n = 1,2, \cdots. 
\label{eq_op_end}
\end{align}
These relations are immediately derived from the definitions.

\section{Two examples of the formulation}
\label{sec_examples}

\subsection{Canonical commutation relation}
\label{sec_ccr}

The interacting Fock space associated with a Jacobi sequence $\{\omega_n = n\}$
is called the \textit{Boson Fock space} \cite{Hora_Obata_book}.
In this case, we obtain the following \textit{canonical commutation relation} (CCR):
\begin{align}
B^- B^+ - B^+ B^- = 1.
\end{align}
It has been shown that the CCR is useful in order to investigate chemical reaction systems,
particle creation-annihilation systems, and ecological systems.
We here note that the scheme based on the CCR corresponds to the Doi-Peliti formalism.

In Ref.~\onlinecite{Ohkubo2012}, it has been suggested that 
the Doi-Peliti formalism is related to orthogonal polynomials,
i.e., the Hermite polynomials and the Charlier polynomials.
Using the scheme introduced in the present paper,
this connection becomes more clearly, as follows.

Let $\mathfrak{P}_\mathrm{fm}(\mathbb{R})$ 
be the set of probability measures on $\mathbb{R}$ having
finite moments of all orders.
Let $\{\Xi_n(x)\}$ be the orthogonal polynomials associated with
$\mu \in \mathfrak{P}_\mathrm{fm}(\mathbb{R})$.
From the theory of orthogonal polynomials, 
there exists a pair of sequences $\alpha_1, \alpha_2, \dots \in \mathbb{R}$
and $\omega_1, \omega_2, \dots > 0$ uniquely
determined by
\begin{align}
&\Xi_0(x) = 1, \quad \Xi_1(x) = x - \alpha_1, \label{eq_three_term_1} \\
& x \Xi_n(x) = \Xi_{n+1}(x) + \alpha_{n+1} \Xi_n(x) + \omega_n \Xi_{n-1}(x), \quad n = 1,2,\cdots.
\label{eq_three_term_2}
\end{align}
For example, 
we have $\alpha_n \equiv 0$ and $\omega_n = n$ for the Hermite polynomials;
for the Charlier polynomials,
$\alpha_n = \lambda + n-1$ and $\omega_n = \lambda n$,
where $\lambda > 0$.
Although some modifications are needed for the Charlier polynomials
in order to remove the effect of $\lambda$,
these two orthogonal polynomials have essentially the same form of $\{\omega_n\}$ (the same $n$ dependency).
It is clear that the above $\{\omega_n\}$ is a Jacobi sequence.
This means that we have, in principle, various orthogonal polynomials 
in order to describe the CCR by choosing the sequence $\{\alpha_n\}$.

In order to discuss a general case,
we here employ $\omega_n = \lambda n$.
In this case, we have a slightly different form of commutation relation,
i.e., $B^- B^+ - B^+ B^- = \lambda$.
This commutation relation is rewritten as 
$(B^- / \lambda) B^+ - B^+ (B^-/\lambda) = 1$,
and we recover the CCR by defining $B^-/\lambda$ as a new operator: $B_{(\lambda)}^- \equiv B^-/\lambda$.
According to this new definition,
the actions of $B_{(\lambda)}^+ \equiv B^+$ and $B_{(\lambda)}^-$ on the bra and ket vectors
are written as follows:
\begin{align}
&B_{(\lambda)}^+ | n \rangle = | n+1 \rangle, \quad n = 0,1,2,\cdots, \label{eq_new_op_start}\\
&B_{(\lambda)}^- | 0 \rangle = 0, \qquad B_{(\lambda)}^- | n \rangle = n | n-1 \rangle, \quad n = 1,2,\cdots,\\
&\langle n | B_{(\lambda)}^- = \langle n+1 | \lambda^{-1}, \quad n = 0,1,2, \cdots, \\
&\langle 0 | B_{(\lambda)}^+ = 0, \qquad \langle n | B_{(\lambda)}^+ = \langle n-1 | n \lambda, 
\quad n = 1,2, \cdots,
\label{eq_new_op_end}
\end{align}
which have already been introduced in Ref.~\onlinecite{Ohkubo2012}
as a one-parameter extension of the Doi-Peliti formalism.

The explicit constructions for the one-parameter extension
has already been given in Ref.~\onlinecite{Ohkubo2012}.
We here review a construction based on the Hermite polynomials.
When we take $\alpha_n \equiv 0$ and $\omega_n = \lambda n$,
the following rescaled Hermite polynomials $\{\tilde{H}^{(\lambda)}_n(x)\}$ 
satisfies Eqs.~\eqref{eq_three_term_1} and \eqref{eq_three_term_2}:
\begin{align}
\tilde{H}^{(\lambda)}_n(x) \equiv \sqrt{ \left( \frac{\lambda}{2} \right)^n} 
H_n\left( \frac{x}{\sqrt{2\lambda}}\right),
\end{align}
where
\begin{align}
H_n(x) = (-1)^n \rme^{x^2} \frac{\rmd^n}{\rmd x^n} \rme^{-x^2}
\end{align}
is the usual Hermite polynomials.
The rescaled Hermite polynomials satisfy the following orthogonality relation:
\begin{align}
\int_{-\infty}^{+\infty} \tilde{H}^{(\lambda)}_n(x) \tilde{H}^{(\lambda)}_m(x) 
\mu^{(\lambda)}(x) \rmd x
= \lambda^n n! \, \delta_{n,m},
\label{eq_Hermite_orthogonality}
\end{align}
where 
\begin{align}
\mu^{(\lambda)}(x) = \frac{1}{\sqrt{2\pi \lambda}} \rme^{- x^2 / (2\lambda)}.
\end{align}
Hence, we define the ket vectors, the bra vectors, and the linear operators $B^\pm$ as follows:
\begin{align}
&| n \rangle \equiv \tilde{H}^{(\lambda)}_n(x), \qquad
\langle n | \equiv \int_{-\infty}^{\infty} \rmd x \mu^{(\lambda)}(x) \tilde{H}^{(\lambda)}_n(x) (\cdot), \\
&B_{(\lambda)}^+ \equiv x - \lambda \frac{\rmd}{\rmd x}, \qquad
B_{(\lambda)}^- \equiv \frac{\rmd}{\rmd x}.
\end{align}
Note that the integration in the bra vector $\langle n |$ should be considered
after taking the inner product with the ket vector $| n \rangle$;
we indicate this fact using $(\cdot)$.
We can confirm that the above definitions satisfy the properties
in Eqs.~\eqref{eq_new_op_start}-\eqref{eq_new_op_end} adequately.
In addition, the projection state $\langle \mathrm{P} |$ can be expressed as
\begin{align}
\langle \mathrm{P} | = \langle 0 | \exp(B_{(\lambda)}^-) = \sum_{n=0}^\infty \frac{1}{\lambda^n n!} \langle n |,
\end{align}
which is consistent with the previous work \cite{Ohkubo2012}.

Why is the CCR useful? 
One of the reasons is that transition rates in certain types of stochastic processes
have a specific dependency on the state $n$.
Here, we use a simple birth-coagulation process as an example.
The birth reaction $X \to X + X$ occurs with the rate constant $\alpha$
for each particle, and its backward reaction (coagulation) $X+X \to X$ occurs with $\beta$.
The master equation for the process is written as
\begin{align}
\frac{\rmd}{\rmd t} P(n,t) = \alpha [(n-1)P(n-1,t) - n P(n,t)] + \beta [n (n+1) P(n+1,t) - n(n-1) P(n,t)],
\label{eq_master}
\end{align}
where $P(n,t)$ is a probability of finding $n$ particles at time $t$,
and we define $P(-1,t)\equiv 0$.
As one can see, the dependency of the transition rates on the number of particles $n$
seems to be easily dealt with the formalism in Eqs.~\eqref{eq_new_op_start}-\eqref{eq_new_op_end}.
(For simplicity, we here use $\lambda = 1$ and describe $B_{(\lambda)}^\pm$ as $B^\pm$.)
Actually, the time evolution equation 
for a state vector defined as Eq.~\eqref{eq_state_vector}
is written as
\begin{align}
\frac{\rmd}{\rmd t} | \psi(t) \rangle 
= L(B^+,B^-) | \psi(t) \rangle,
\label{eq_Schrodinger_like}
\end{align}
where 
\begin{align}
L(B^+,B^-) =
\alpha (B^+ - 1) B^+ B^-
+ \beta (1 - B^+) 
B^+ (B^-)^2.
\label{eq_operator_example}
\end{align}
This simple expression of the time evolution operator $L(B^+,B^-)$ 
enables us to investigate the system in various analytical methods
developed in quantum mechanics,
e.g.,
a coherent-state path-integral formulation,
perturbation calculations,
renormalization group method, and so on.


We finally comment on some relations with previous works for the Doi-Peliti formalism.
For the Doi-Peliti formalism,
the following expressions for the bra and ket vectors have been known \cite{Droz1995}:
\begin{align}
| n \rangle \equiv x^n, \qquad
\langle m | \equiv \int \mathrm{d} x \, 
\delta(x) \left( \frac{\mathrm{d}}{\mathrm{d} x} \right)^m (\cdot),
\end{align}
where $\delta(x)$ is the Dirac delta function.
The corresponding linear operators $B^\pm$ are as follows:
\begin{align}
B^+ \equiv x, \qquad B^- \equiv \frac{\mathrm{d}}{\mathrm{d} x}.
\end{align}
These vectors and operators satisfy Eqs.~\eqref{eq_new_op_start}-\eqref{eq_new_op_end}
with $\lambda = 1$.
This construction is based on a generating function approach,
and even in this case, 
the definition of the state $\varphi$ in Eq.~\eqref{eq_new_definition_of_state} 
is useful and available.



\subsection{Canonical anticommutation relation}
\label{sec_car}

Here, we consider a simple hopping model,
which has been used to study pumping phenomena in nonequilibrium physics \cite{Sinitsyn2007}.
The system consists of three parts,
i.e., a left particle reservoir, a right particle reservoir, and a container.
The left and right particle reservoirs are assumed to be large enough,
and we only consider a change of state of the container.
The container can have at most one particle,
so that the state of the container is `empty' or `filled'. 
When the container is filled with one particle
the particle can escape from the container by jumping into one of the two particle reservoirs.
The transition rates, $k_1, k_{-1}, k_2, k_{-2}$, are defined as the following scheme:
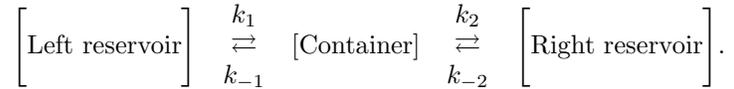
\begin{align*}
\Bigg[ \textrm{Left reservoir} \Bigg] 
\quad
\begin{matrix}
k_1 \\ \rightleftarrows \\ k_{-1}
\end{matrix} 
\quad
[ \mathrm{Container} ] 
\quad
\begin{matrix}
k_2 \\ \rightleftarrows \\ k_{-2}
\end{matrix}
\quad
\Bigg[ \textrm{Right reservoir} \Bigg]. 
\end{align*}
We define $p_\mathrm{e}(t)$ ($p_\mathrm{f}(t)$) as 
the probability with which the container is empty (filled)
at time $t$.
The master equation for the time evolution of $p_\mathrm{e}(t)$ and $p_\mathrm{f}(t)$ is
\begin{align}
\frac{\rmd}{\rmd t} 
\begin{pmatrix}
p_\mathrm{f}(t) \\ p_\mathrm{e}(t)
\end{pmatrix}
= 
\begin{pmatrix}
-k_{-1} - k_2 & k_1 + k_{-2} \\
k_{-1} + k_2 & -k_1 - k_{-2}
\end{pmatrix}
\begin{pmatrix}
p_\mathrm{f}(t) \\ p_\mathrm{e}(t)
\end{pmatrix}.
\label{eq_master_hopping}
\end{align}

In order to discuss the hopping model,
the following Jacobi sequence is useful:
$\omega_1 = 1$ and $\omega_n = 0$ for $n = 2,3,\cdots$.
In this case, we have \textit{canonical anticommutation relation}:
\begin{align}
B^- B^+ + B^+ B^- = 1.
\end{align}
For clarity, the creation and annihilation operators for the 
canonical anticommutation relation are denoted as 
$\sigma^+ \equiv B^+$ and  $\sigma^- \equiv B^-$, respectively.
The following definitions are useful:
\begin{align}
e_0 \equiv \begin{pmatrix} 0 \\ 1 \end{pmatrix}, \quad
e_1 \equiv \begin{pmatrix} 1 \\ 0 \end{pmatrix},
\end{align}
\begin{align}
\sigma^+ \equiv \begin{pmatrix}
0 & 1 \\
0 & 0
\end{pmatrix}, \quad
\sigma^- \equiv \begin{pmatrix}
0 & 0 \\
1 & 0
\end{pmatrix}.
\end{align}
The actions of $\sigma^{\pm}$ to the vectors $e_0$ and $e_1$ are as follows:
\begin{align*}
e_0 \xrightarrow{\sigma^+} e_1 \xrightarrow{\sigma^+} 0,
\qquad e_1 \xrightarrow{\sigma^-} e_0 \xrightarrow{\sigma^-} 0.
\end{align*}
Using the above vectors and operators,
we have the corresponding time evolution equation for 
the state vector $| \psi(t) \rangle$, which is defined as
\begin{align}
| \psi(t) \rangle = p_\mathrm{e}(t) e_0 + p_\mathrm{f}(t) e_1,
\end{align}
as follows:
\begin{align}
\frac{\rmd}{\rmd t} | \psi(t) \rangle &= L(\sigma^+,\sigma^-) | \psi(t) \rangle, \label{eq_te_hopping}
\end{align}
where
\begin{align}
L(\sigma^+,\sigma^-) &= (-k_{-1}-k_2) \sigma^+ \sigma^- + (k_1 + k_{-2}) \sigma^+
+ (k_{-1} + k_2) \sigma^- + (-k_1-k_{-2}) \sigma^- \sigma^+ 
\label{eq_op_hopping}.
\end{align}
Note that 
\begin{align}
\sigma^- \sigma^+ = \begin{pmatrix}
0 & 0 \\
0 & 1
\end{pmatrix}, \quad
\sigma^+ \sigma^- = \begin{pmatrix}
1 & 0 \\
0 & 0
\end{pmatrix}.
\end{align}
It is sometimes useful to analyze Eq.~\eqref{eq_te_hopping} instead of Eq.~\eqref{eq_master_hopping};
for example, 
Eqs.~\eqref{eq_te_hopping} and \eqref{eq_op_hopping} is useful
in order to discuss the counting statistics in the following section.

We finally note an extension for multiple state cases.
The hopping model can be formulated based on $M(2,\mathbb{C})$,
because there are only two possible states (empty or filled).
It is an easy task to extend it to $n$-state cases,
which are formulated in terms of $M(n,\mathbb{C})$.
Hence, in principle, it is possible to describe
a stochastic process with a finite number of states
in terms of the algebraic probability.

\section{Application to counting statistics}
\label{sec_cs}

\subsection{Aims of counting statistics}
\label{sec_cs_aim}

As an applications of the formalism of the algebraic probability,
we here derive a scheme in the counting statistics \cite{Gopich2003,Gopich2005,Gopich2006}.
In the counting statistics, we count the number of specific transitions.
For example, 
we here count the number of hopping between the container and the right reservoir
in the hopping model in Sec.~\ref{sec_car}.
Using the counting statistics,
all statistics, including the higher-order moment, can be evaluated.
A stochastic process is described by a transition matrix,
and it has been shown that
a slightly modified transition matrix plays an important role
in the counting statistics.
While a derivation based on a generating function
has already been given,
we will show that the scheme in the counting statistics
can be treated within the algebraic probability;
the modified transition matrix is obtained from an intuitive discussion.

\subsection{Basic equations in counting statistics}

We firstly explain basic equations used in the counting statistics.
In order to explain them, we use the hopping model introduced in Sec.~\ref{sec_car}.
(The hopping model is a nice toy model 
in order to investigate nonequilibrium behavior, and 
actually the counting statistics for the hopping model has been already studied 
\cite{Sinitsyn2007,Ohkubo2007,Ohkubo2008}.)

A net flow between two transitions is often investigated in nonequilibrium physics.
For example, in the hopping model,
the net flow from the container to the right reservoir is calculated
as a subtraction of 
``(i) the number of hopping from the right reservoir to the container''
from 
``(ii) the number of hopping from the container to the right reservoir''. 

In order to investigate the statistics of the net flow,
it is a common way to construct a generating function.
Let $P(N_\mathrm{A}|t)$ be the probability with which 
there are $N_\mathrm{A}$ net transitions 
from the container to the right reservoir during time $t$.
Since $N_\mathrm{A}$ is defined as the subtraction of (i) from (ii),
$N_\mathrm{A}$ is an integer value and it can be negative.
The generating function $Z(\phi,t)$ for the statistics is defined as follows:
\begin{align}
Z(\phi,t) = \sum_{N_\mathrm{A}=-\infty}^\infty P(N_\mathrm{A}|t) \rme^{\rmi N_\mathrm{A} \phi}.
\end{align}
Using the generating function $Z(\phi,t)$, all statistics related to the net flow
can be evaluated.
However, at this stage, the probability $P(N_\mathrm{A}|t)$ is not known.
In stead of an explicit calculation of $P(N_\mathrm{A}|t)$,
it has been shown that the following basic equation is useful \cite{Sinitsyn2007}:
\begin{align}
\frac{\rmd}{\rmd t} 
\begin{pmatrix}
f_\mathrm{f}(\phi, t) \\ f_\mathrm{e}(\phi, t)
\end{pmatrix}
= 
\begin{pmatrix}
-k_{-1} - k_2 & k_1 + k_{-2} \rme^{- \rmi \phi} \\
k_{-1} + k_2 \rme^{\rmi \phi}& -k_1 - k_{-2}
\end{pmatrix}
\begin{pmatrix}
f_\mathrm{f}(\phi,t) \\ f_\mathrm{e}(\phi,t)
\end{pmatrix}.
\label{eq_cs_basic}
\end{align}
Using the solution of Eq.~\eqref{eq_cs_basic},
the generating function is evaluated as
\begin{align}
Z(\phi,t) = f_\mathrm{f}(\phi,t) + f_\mathrm{e}(\phi,t).
\end{align}
Note that the matrix in Eq.~\eqref{eq_cs_basic} is very similar to the transition matrix
of the original master equation (Eq.~\eqref{eq_master_hopping}),
but this matrix does not satisfy the probability conservation.
We here call it the \textit{modified transition matrix}.

The derivation of the basic equation (Eq.~\eqref{eq_cs_basic}) has been performed
in terms of the generating function approach.
(For example, see Ref.~\onlinecite{Gopich2005}.)
However, the derivation is a little complicated,
especially when one wants to count multi-target transitions, like the example here;
we here count not only (i), but also (ii) simultaneously.

In the following subsections, we explain how the basic equation is easily derived
from the discussions of the algebraic probability.

\subsection{Combination of two interacting Fock space}
\label{sec_cs_two_Fock}

In the scheme of the algebraic probability discussed in Sec.~\ref{sec_algebraic_probability},
the index $n$ cannot be negative.
However, as explained before, 
the net flow can be negative, and hence the negative variables should be considered.
In order to treat the negative $n$ for the counting statistics,
we here combine two interacting Fock space;
one describes a positive contribution
and another corresponds to a negative contribution.

While there may be various constructions for the two interacting Fock space,
we here employ two spaces with the CCR  (with $\lambda = 1$).
Hence, a joint probability distribution is used, i.e., 
$P(n_\mathrm{p},n_\mathrm{n})$ for
$n_\mathrm{p} = 0,1,2, \cdots$ and $n_\mathrm{n} = 0,1,2, \cdots$.
The variable $n_\mathrm{p}$ corresponds to the number of transitions
corresponding to the positive contribution,
and $n_\mathrm{n}$ to the negative contribution.
Using the following ket vector
\begin{align}
| n_\mathrm{p}, n_\mathrm{n} \rangle = | n_\mathrm{p} \rangle \otimes | n_\mathrm{n} \rangle,
\end{align}
we define the linear operators $B_\mathrm{p}^{\pm}$ as follows:
\begin{align}
B_\mathrm{p}^+ = B^+ \otimes \mathbf{1}, \qquad B_\mathrm{p}^- = B^- \otimes \mathbf{1},
\qquad B_\mathrm{n}^+ = \mathbf{1} \otimes B^+, \quad B_\mathrm{n}^- = \mathbf{1} \otimes B^-.
\end{align}
For example, the creation and annihilation operators for the positive contribution part,
$B_\mathrm{p}^\pm$, act on the ket vector $| n_\mathrm{p}, n_\mathrm{n}\rangle$ as follows:
\begin{align}
& B_\mathrm{p}^+ | n_\mathrm{p}, n_\mathrm{n} \rangle = | n_\mathrm{p}+1, n_\mathrm{n} \rangle, 
\quad B_\mathrm{p}^- | n_\mathrm{p}, n_\mathrm{n} \rangle = n_\mathrm{p} | n_\mathrm{p}-1, n_\mathrm{n} \rangle, \\
&N_\mathrm{p} | n_\mathrm{p}, n_\mathrm{n} \rangle 
= B_\mathrm{p}^+ B_\mathrm{p}^- | n_\mathrm{p}, n_\mathrm{n} \rangle
= n_\mathrm{p} | n_\mathrm{p}, n_\mathrm{n} \rangle,
\end{align}
for all $n_\mathrm{p}$.
The inner product between the bra and ket vectors becomes
\begin{align}
\langle m_\mathrm{p}, m_\mathrm{n} | n_\mathrm{p}, n_\mathrm{n} \rangle = 
n_\mathrm{p}! n_\mathrm{n}! \delta_{m_\mathrm{p},n_\mathrm{p}} \delta_{m_\mathrm{n},n_\mathrm{n}}.
\end{align}
For the negative contribution case,
two linear operators $B_\mathrm{n}^{\pm}$ are defined in a similar manner.
Next, we consider analytic representations for the bra and ket vectors,
and the linear operators $B_{\mathrm{p,n}}^{\pm}$.
In order to connect the discussions below
with the generating function approach,
we use the following definition of the Kronecker delta:
\begin{align}
\delta_{m,n} = \frac{1}{2\pi} \int_0^{2\pi} \rmd \phi \, \rme^{\rmi (n-m)\phi}.
\end{align}
Hence, the following analytical representations are obtained:
\begin{align}
&| n_\mathrm{p} \rangle = \rme^{\rmi n_\mathrm{p} \phi_\mathrm{p}},
\qquad \langle m_\mathrm{p} | 
= \frac{m_\mathrm{p}!}{2\pi} \int_0^{2\pi} \rmd \phi_\mathrm{p} \, \rme^{-\rmi m_\mathrm{p} \phi_\mathrm{p}} (\cdot), \\
&B_\mathrm{p}^+ = \rme^{\rmi \phi_\mathrm{p}}, 
\qquad B_\mathrm{p}^- = - \rmi \rme^{- \rmi \phi_\mathrm{p}} \frac{\rmd}{\rmd \phi_\mathrm{p}}, 
\qquad N_\mathrm{p} = B_\mathrm{p}^+ B_\mathrm{p}^- = - \rmi \frac{\rmd}{\rmd \phi_\mathrm{p}}.
\end{align}
For the negative contribution case, similar representations are obtained.
From the construction, we have $N_\mathrm{p} | n_\mathrm{p}, n_\mathrm{n} \rangle \ge 0$
and $N_\mathrm{n} | n_\mathrm{p}, n_\mathrm{n} \rangle \ge 0$;
these schemes do not violate properties of the algebraic probability.

In order to count the net flow,
we must consider the operator $N_\mathrm{p} - N_\mathrm{n}$,
which means that $N_\mathrm{p}$ gives the positive contribution
and $N_\mathrm{n}$ describes the negative contribution.
Hence, the following number operator should be introduced:
\begin{align}
N_\mathrm{c} \equiv B_\mathrm{p}^+ B_\mathrm{p}^- - B_\mathrm{n}^+ B_\mathrm{n}^-.
\end{align}
Next, we employ variable transformations:
$\phi_\mathrm{p} = \phi$, and $\phi_\mathrm{n} = \phi' - \phi$.
Due to the variable transformations,
the ket vector $|n_\mathrm{p}, n_\mathrm{n} \rangle$ becomes
\begin{align}
| n_\mathrm{p}, n_\mathrm{n} \rangle = \rme^{\rmi \phi_\mathrm{p}} \rme^{\rmi \phi_\mathrm{n}}
= \rme^{\rmi (n_\mathrm{p} - n_\mathrm{n}) \phi + \rmi n_\mathrm{n} \phi'},
\end{align}
and we have
\begin{align}
N_\mathrm{c} = - \rmi \frac{\partial}{\partial \phi}.
\end{align}
Actually, the number operator $N_\mathrm{c}$ acts on $|n_\mathrm{p}, n_\mathrm{n} \rangle$
as
\begin{align}
N_\mathrm{c} |n_\mathrm{p}, n_\mathrm{n} \rangle 
= (n_\mathrm{p} - n_\mathrm{n}) |n_\mathrm{p}, n_\mathrm{n} \rangle.
\end{align}

Notice that the operators $B_\mathrm{p}^-$ or $B_\mathrm{n}^-$ does not used alone
in the counting statistics:
$n_\mathrm{p}$ and $n_\mathrm{n}$ cannot decrease with time,
and there is no need to consider the single action of $B_\mathrm{p}^-$ and $B_\mathrm{n}^-$ in the formalism.
If we want to decrease the number of transitions,
we should increase $n_\mathrm{n}$ instead of the decrease of $n_\mathrm{p}$.
This indicates that we can set $\phi' = 0$ for simplicity, 
and the following two operators are introduced:
\begin{align}
B_\mathrm{c}^+ \equiv \rme^{\rmi \phi}, 
\quad B_\mathrm{c}^- \equiv \rme^{- \rmi \phi}.
\label{eq_operators_in_counting_statistics}
\end{align}
In addition, introducing $n_\mathrm{c} \in \mathbb{Z}$ and $n_\mathrm{c} = n_\mathrm{p} - n_\mathrm{n}$,
the following ket vector, which stems from $| n_\mathrm{p}, n_\mathrm{n} \rangle$, 
should be useful in the counting statistics:
\begin{align}
| n_\mathrm{c} \rangle = \rme^{\rmi n_\mathrm{c} \phi}.
\end{align}
In summary, we have the following relations:
\begin{align}
&B_\mathrm{c}^+ | n_\mathrm{c} \rangle = | n_\mathrm{c}+1 \rangle, 
\quad B_\mathrm{c}^- | n_\mathrm{c} \rangle = | n_\mathrm{c}-1 \rangle, \\
&N_\mathrm{c} | n_\mathrm{c} \rangle = n_\mathrm{c} | n_\mathrm{c} \rangle.
\end{align}
The above constructions means that
the ket vectors suitable for the counting statistics
can be generated adequately from the two interacting Fock spaces.
Notice that this scheme does not violate the properties of the algebraic probability.
While $n_\mathrm{c}$ can take a negative value,
the positivity of the state ($\varphi(a^*a) \ge 0$) is still satisfied
when we consider a $*$-algebra is constructed by $B_\mathrm{c}^{\pm}$ and $N_\mathrm{c}$.
(Note that $(N_\mathrm{c})^* = N_\mathrm{c}$.)


\subsection{Derivation of the basic equations in the counting statistics}
\label{sec_cs_main}

We here discuss how the constructed Fock space in Sec.~\ref{sec_cs_two_Fock}
is used in the counting statistics.
Again, as an example, a counting problem for the hopping model 
in Sec.~\ref{sec_car} is considered.

We calculate a net flow from the container to the right reservoir.
If a particle hops from the container to the right reservoir, we count it as $+1$;
$-1$ for a particle hopping from the right reservoir to the container.
This means that the counting process is also a stochastic process;
not only the stochastic process for the hopping model,
but also that for counting statistics should be considered simultaneously.
Hence, we define a joint probability $P(s,n_\mathrm{c},t)$ with which
the system is in $s$ ($s \in \{\mathrm{e},\mathrm{f}\}$) and 
the net flow with $n_\mathrm{c} \in \mathbb{Z}$ at time $t$.
The state vector $| \phi(t) \rangle$ in the algebraic probability is defined as
\begin{align}
| \psi(t) \rangle = \sum_{s\in \{\mathrm{e}, \mathrm{f}\}}
\sum_{n_\mathrm{c}=-\infty}^\infty P(s,n_\mathrm{c},t)
| s, n_\mathrm{c} \rangle.
\end{align}
For the hopping model,
a quantity at time $t$ is evaluated by using the operator $\exp(L(\sigma^+,\sigma^-)t)$,
where $L(\sigma^+,\sigma^-)$ is defined as Eq.~\eqref{eq_op_hopping}.
How should we construct the operator $L_\mathrm{c}(\sigma^+,\sigma^-,B_\mathrm{c}^+,B_\mathrm{c}^-)$
for the counting statistics case?
In order to obtain it, 
we consider small time interval $\Delta t$
and the operator 
$\exp(L(\sigma^+,\sigma^-) \Delta t)$
for the hopping model.
We have
\begin{align}
\exp\left( L(\sigma^+,\sigma^-) \Delta t \right)
\simeq& \bm{1} + L(\sigma^+,\sigma^-) \Delta t \nonumber \\
=& \left[ \bm{1} + (-k_{-1} - k_2) \Delta t \sigma^+ \sigma^- 
+ (-k_1-k_{-2}) \Delta t \sigma^- \sigma^+ 
\right] \nonumber \\
&+ k_1 \Delta t \sigma^+ + k_{-2} \Delta t \sigma^+  
+ k_{-1} \Delta t \sigma^- + k_2 \Delta t \sigma^- .
\label{eq_exp_op_cs}
\end{align}
Each term in Eq.~\eqref{eq_exp_op_cs} can be interpreted as follows.
The second term means that 
there is a state change caused by $\sigma^+$
and its probability is $k_1 \Delta t$.
With this probability, 
a particle is hopping from the left reservoir to the container.
The third term
corresponds to a particle hopping 
from the right reservoir to the container,
and so on.
The first term shows a case in which no transition occurs.
In our problem settings,
the net flow from the container to the right reservoir
should be calculated,
and hence we immediately notice that 
the following linear operator $L_\mathrm{c}(\sigma^+,\sigma^-, B_\mathrm{c}^+,B_\mathrm{c}^-) $ 
for the time evolution of $| \psi \rangle$ should be used:
\begin{align}
&L_\mathrm{c}(\sigma^+,\sigma^-, B_\mathrm{c}^+,B_\mathrm{c}^-) \nonumber \\
&\quad = (-k_{-1}-k_2) \sigma^+ \sigma^- + (k_1 + k_{-2} B_\mathrm{c}^-) \sigma^+
+ (k_{-1} + k_2 B_\mathrm{c}^+) \sigma^- + (-k_1-k_{-2}) \sigma^- \sigma^+ .
\end{align}
That is, the transition from the container to the right reservoir 
(the term $k_{2}\sigma^-$) is followed by $B_\mathrm{c}^+$.
In addition, $B_\mathrm{c}^-$ is combined
with the transition from the right reservoir to the container 
(the term $k_{-2} \sigma^+$).
In a matrix formulation,
we obtain
\begin{align}
L_\mathrm{c}(\sigma^+,\sigma^-, B_\mathrm{c}^+,B_\mathrm{c}^-) 
= \begin{pmatrix}
-k_{-1} - k_2 & k_1 + k_{-2} \rme^{-\rmi \phi}\\
k_{-1} + k_2 \rme^{\rmi \phi} & -k_1 - k_{-2}
\end{pmatrix},
\end{align}
where we used Eq.~\eqref{eq_operators_in_counting_statistics}.
This is actually the modified transition matrix in Eq.~\eqref{eq_cs_basic},
and as a result,
the above discussions give the basic equation for counting statistics
immediately.
As exemplified here,
the operator or the modified transition matrix for the counting statistics
is easily and intuitively obtained using the scheme of the algebraic probability.

\subsection{Counting statistics for continuous values}
\label{sec_cs_continuous}

We sometimes need to consider cases with non-integer counting.
For example, when one wants to calculate entropy productions in a stochastic process,
each transition causes entropy production,
and the entropy production is not an integer but a continuous value.
We will show that the construction based on the algebraic probability
is also available for the non-integer cases.

We assume that a certain transition occurs and
an quantity $\eta$ is produced due to the transition.
Note that $\eta$ is a continuous value and it can be negative in general.
In order to apply the counting statistics explained in Sec.~\ref{sec_cs_main},
let introduce a small interval $\Delta$ and 
approximate $\eta$ as follows:
\begin{align}
\eta = n_\eta \Delta + \varepsilon, \quad \varepsilon \ge 0,
\end{align}
where
$n_\eta = \lfloor \eta / \Delta \rfloor$
and we used the floor function
\begin{align}
\lfloor \eta \rfloor = \max\{ n \in \mathbb{Z} | n \le \eta\}.
\end{align}
Since $n_\eta \Delta$ is an integer,
we can use the scheme of the counting statistics explained above.
In the counting statistics, there is only one increment $+1$ 
(or one decrement $-1$) due to a target transition,
so that we employ $B_\mathrm{c}^+$ (or $B_\mathrm{c}^-$).
In the non-integer case, one transition occurs the change with $\eta$,
and hence we should use $(B_\mathrm{c}^+)^{n_\eta \Delta}$.
(If $n_\eta < 0$, this operator corresponds to 
$(B_\mathrm{c}^-)^{\lvert n_\eta \rvert \Delta}$ due to Eq.~\eqref{eq_operators_in_counting_statistics}, 
so that there is no need to consider $B_\mathrm{c}^-$ explicitly here.)
Taking the limit $\Delta \to 0$,
we obtain $(B_\mathrm{c}^+)^{n_\eta \Delta} \simeq \rme^{\rmi \eta \phi}$,
which is consistent with the previous work
for the entropy production \cite{Sagawa2011}.

\section{Concluding remarks}
\label{sec_conclusions}

In the present paper, we gave a definition of a state
in algebraic probability,
which is suitable for classical stochastic processes.
Based on a Jacobi sequence, an interacting Fock space is constructed.
Using this construction, 
the Doi-Peliti formalism can be connected to the orthogonal polynomials naturally.
In addition, the scheme is applied to the counting statistics,
which enables us to understand the meanings of the modified transition matrix
in the generating function approach;
the multiplied factors in the modified transition matrix 
correspond to creation and annihilation operators for the counting process.
As a result, it is clarified that 
we can deal with various methods studied previously in a unified way.

One may consider that
commutative operators correspond to classical systems,
and noncommutative ones mean quantum systems.
However, as clarified in the present paper,
the noncommutative relation does not mean directly quantum systems;
certain types of classical stochastic processes 
are adequately dealt with the noncommutative operators.
Only a difference between a quantum case and a classical stochastic process
is within the usage of the probability $P(n)$, not the square root,
for the definition of the state.



\section*{ACKNOWLEDGMENTS}

This work was supported in part by grant-in-aid for scientific research 
(No. 20115009)
from the Ministry of Education, Culture, Sports, Science and Technology (MEXT), Japan.

\end{document}